\documentclass[11pt]{article}
\usepackage[final]{acl}
\usepackage{times}
\usepackage{latexsym}
\usepackage{amsmath,amssymb}
\usepackage{graphicx}
\usepackage{url}
\usepackage{booktabs}
\usepackage{enumitem}
\usepackage{microtype}
\usepackage{tikz}
\usetikzlibrary{shapes.geometric, arrows.meta, positioning}
\usepackage{pgfplots}
\pgfplotsset{compat=1.18}
\usepackage{wrapfig}
\usepackage{capt-of}

\title{Smart Privacy Policy Assistant: An LLM-Powered System for Transparent and Actionable Privacy Notices}

\author{
\textbf{
Sriharshini Kalvakuntla$^{1}$,
Luoxi Tang$^{1}$,
Yuqiao Meng$^{1}$,
Zhaohan Xi$^{1}$
}\\
$^{1}$Binghamton University\\
\textbf{Correspondence:} \href{mailto:zxi1@binghamton.edu}{zxi1@binghamton.edu}
}

\begin{document}
\maketitle

\begin{abstract}
Most users agree to online privacy policies without reading or understanding them, even though these documents govern how personal data is collected, shared, and monetized. Privacy policies are typically long, legally complex, and difficult for non-experts to interpret. This paper presents the Smart Privacy Policy Assistant, an LLM-powered system that automatically ingests privacy policies, extracts and categorizes key clauses, assigns human-interpretable risk levels, and generates clear, concise explanations. The system is designed for real-time use through browser extensions or mobile interfaces, surfacing contextual warnings before users disclose sensitive information or grant risky permissions. We describe the end-to-end pipeline, including policy ingestion, clause categorization, risk scoring, and explanation generation, and propose an evaluation framework based on clause-level accuracy, policy-level risk agreement, and user comprehension.
\end{abstract}

\section{Introduction}

\paragraph{Background and Current Status}
Privacy policies are the primary mechanism through which online services disclose their data collection, sharing, and usage practices to users. These documents govern critical aspects of users’ digital lives, including what personal information is collected, processed, and shared. Despite their importance, a large body of empirical research consistently shows that users rarely read or understand privacy policies in practice. McDonald and Cranor \citep{mcdonald2008cost} estimate that reading the privacy policies encountered by an average user would require an infeasible amount of time each year, while Obar and Oeldorf-Hirsch \citep{obar2018biggest} demonstrate that most users consent to policies without engaging with their content at all.

To mitigate this gap, researchers have proposed usability-focused interventions such as layered notices and simplified disclosures \citep{cranor2012necessary, kelley2010standardizing}. Advances in NLP have further enabled large-scale analysis of privacy policies.
 Large Language Models (LLMs), in particular, have shown strong capabilities in processing long documents and generating natural language explanations, creating new opportunities for making privacy disclosures more accessible to non-expert users.

\paragraph{Limitations of Existing Approaches}

\noindent
\begin{minipage}[t]{0.62\columnwidth}
\vspace{0pt}
Despite substantial progress, existing approaches remain insufficient for supporting informed and actionable user decisions. Early automated systems relied on rule-based extraction and keyword matching, which struggled with linguistic variability and legal ambiguity \citep{reidenberg2015disagreeable}.
\end{minipage}\hfill
\begin{minipage}[t]{0.34\columnwidth}
\vspace{0pt}
\centering
\setlength{\abovecaptionskip}{1pt}
\includegraphics[width=\linewidth]{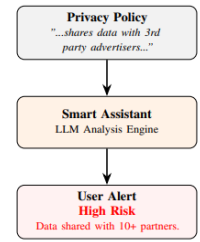}
\vspace{-4pt}
\captionof{figure}{System overview.}
\label{fig:system_overview}
\vspace{-6pt}
\end{minipage}

 Later work introduced annotated privacy policy corpora, such as OPP-115 \citep{wilson2016creation}, enabling supervised classification of policy segments related to data collection, sharing, and user choice; however, these systems often produce fragmented outputs that are difficult for non-expert users to interpret holistically. More recent neural and LLM-based methods improve scalability and fluency, but unconstrained summarization risks omitting critical details, masking harmful practices, or generating explanations that are insufficiently grounded in specific policy clauses. Moreover, most existing approaches provide static summaries rather than context-aware guidance at the moment users are asked to disclose sensitive information or grant permissions, leaving users without clear signals about the relative privacy risk of their decisions.

\paragraph{Overview of Our Approach}
To address these limitations, we propose the \emph{Smart Privacy Policy Assistant}, an LLM-powered system designed to transform privacy policies into structured, interpretable, and actionable guidance. Rather than treating privacy policies as monolithic texts, our system decomposes policies into semantically coherent clauses and maps each clause onto a structured privacy schema capturing data types, collection contexts, sharing recipients, retention practices, tracking technologies, and user controls. This design enables systematic reasoning over privacy practices while preserving traceability to the original policy text.

A central design goal of our system is interpretability. Instead of producing opaque summaries or binary compliance labels, the assistant computes human-interpretable risk scores based on explicitly defined privacy-relevant features. These scores are designed to be monotonic with respect to clearly harmful practices, ensuring that the detection of additional high-risk behaviors cannot reduce the overall assessed risk.

\paragraph{Actionable Explanations and Real-Time Use}
Beyond static analysis, the Smart Privacy Policy Assistant is designed for real-time deployment through browser extensions and mobile interfaces. The system generates concise, clause-grounded explanations that translate legal language into plain descriptions of practical consequences, such as third-party tracking, profiling, or long-term data retention. When users encounter permission requests or data entry forms on high-risk services, the assistant surfaces contextual warnings linked to relevant policy clauses.

\paragraph{Contributions}
Our contributions include an interpretable LLM-based privacy policy analysis system, a monotonic risk scoring framework, and a user-centered evaluation of actionable privacy assistance.

\section{Related Work}

Prior work on privacy policy understanding spans both usable privacy research and automated text analysis. We organize the most relevant literature into two groups.

\paragraph{Usability and Presentation of Privacy Notices}
A large body of work has shown that users rarely read or understand privacy policies, motivating efforts to improve their presentation and usability. Empirical studies demonstrate that the length and complexity of privacy policies make informed consent impractical in real-world settings \citep{mcdonald2008cost, obar2018biggest}. To address this gap, researchers have proposed layered notices, standardized icons, and privacy nutrition labels designed to make disclosures more accessible to users \citep{cranor2012necessary, kelley2010standardizing}. 

\paragraph{Automated Analysis and Language-Based Approaches}
Another line of work focuses on automating the analysis of privacy policies using computational methods. Early approaches relied on rule-based systems and keyword matching, which struggled with legal ambiguity and linguistic variability \citep{reidenberg2015disagreeable}. Subsequent work introduced annotated corpora such as OPP-115, enabling supervised classification of policy segments related to data collection, sharing, retention, and user choice \citep{wilson2016creation}. More recent neural and LLM-based methods explore summarization and simplification of privacy policies, improving scalability and fluency but often producing static summaries that lack explicit grounding, interpretability, or real-time decision support. In contrast to these approaches, our work combines LLM-based analysis with a structured privacy schema and interpretable risk scoring to support actionable, context-aware user guidance.

\section{Method}

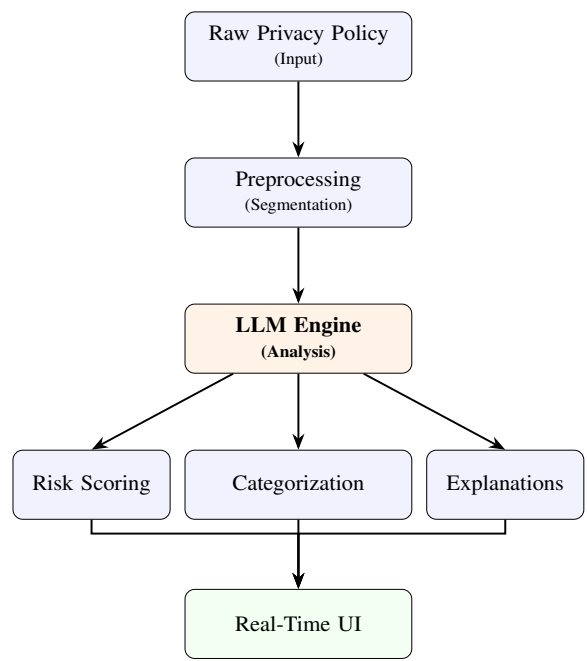
\begin{figure}[ht]
\centering
\resizebox{\columnwidth}{!}{
\begin{tikzpicture}[
    node distance=1.1cm,
    block/.style={rectangle, draw, fill=blue!5, text width=3cm, text centered, rounded corners, minimum height=1cm, font=\small},
    llmblock/.style={rectangle, draw, fill=orange!10, text width=3cm, text centered, rounded corners, minimum height=1cm, font=\small\bfseries},
    arrow/.style={-Stealth, thick}
]

\node (input) [block] {Raw Privacy Policy \\ \scriptsize{(Input)}};
\node (prep) [block, below=of input] {Preprocessing \\ \scriptsize{(Segmentation)}};
\node (llm) [llmblock, below=of prep] {LLM Engine \\ \scriptsize{(Analysis)}};

\node (schema) [block, below=of llm] {Categorization};
\node (risk) [block, left=0.2cm of schema, text width=2cm] {Risk Scoring};
\node (explain) [block, right=0.2cm of schema, text width=2cm] {Explanations};

\node (ui) [block, below=1cm of schema, fill=green!5] {Real-Time UI};

\draw [arrow] (input) -- (prep);
\draw [arrow] (prep) -- (llm);
\draw [arrow] (llm) -- (risk.north);
\draw [arrow] (llm) -- (schema.north);
\draw [arrow] (llm) -- (explain.north);
\draw [arrow] (risk.south) -- ++(0,-0.2) -| (ui.north);
\draw [arrow] (schema.south) -- (ui.north);
\draw [arrow] (explain.south) -- ++(0,-0.2) -| (ui.north);

\end{tikzpicture}
}
\caption{The end-to-end pipeline of our system.}
\label{fig:method_pipeline}
\end{figure}

The Smart Privacy Policy Assistant is designed as an end-to-end system that transforms raw privacy policies into structured representations, interpretable risk assessments, and actionable explanations for end users. The method integrates policy ingestion, clause-level analysis, risk scoring, and explanation generation, and is supported by a curated corpus of real-world privacy policies.

\subsection{Policy Collection and Preprocessing}

\paragraph{Policy Collection}
We collect privacy policies from a diverse set of publicly available online services, including social media platforms, e-commerce websites, VPN providers, fitness and health-tracking applications, and productivity tools. Policies are retrieved by automatically identifying privacy policy links from website markup or by using curated mappings between domains and policy URLs. For mobile applications, policies are obtained from app store listings or developer-provided links.

\paragraph{Text Extraction and Segmentation}
Privacy policies often contain complex HTML structures, navigation elements, and cross-references. The preprocessing pipeline removes boilerplate content, extracts the main policy text while preserving section headers, and normalizes formatting such as whitespace and bullet lists. Policies are then segmented into semantically coherent units, typically corresponding to paragraphs or list items. For PDF-based policies, OCR and layout-aware parsing are applied prior to segmentation.

\subsection{Privacy Schema and Clause Categorization}

To convert unstructured policy text into structured representations, we define a privacy schema that captures core dimensions of data practices, including data types, collection context, sharing recipients, retention and deletion practices, tracking technologies, user controls, and permissions. Each policy segment is analyzed independently.

An LLM is prompted to assign one or more schema categories to each segment, identify relevant data types and actors, and output a structured representation. To reduce unconstrained generation, the model is instructed to select labels from predefined category sets. Outputs are validated against the schema, and segments that cannot be confidently classified are explicitly marked as ambiguous rather than forced into incorrect categories.

\subsection{Risk Scoring}

Each privacy policy is represented as a feature vector encoding the presence or absence of privacy-relevant practices, such as sensitive data collection, third-party data sharing, long retention periods, use of tracking technologies, and availability of user controls.

For a policy $P$, the feature vector $x(P)$ is mapped to a numerical risk score $s \in [0,100]$. The scoring function is designed to be monotonic with respect to clearly harmful practices, ensuring that the detection of additional high-risk behaviors cannot decrease the overall score. For user-facing presentation, the numerical score is discretized into Low, Medium, or High risk categories.

Additional analysis of risk score distributions is provided in Appendix~A.

\subsection{Explanation Generation}

To support user understanding, the system generates concise, targeted explanations aligned with detected risks. Explanations are grounded in specific policy segments and summarize the practical implications of identified practices, such as third-party tracking, profiling, or long-term data retention. The generation process avoids speculative language and explicitly ties explanations to detected clauses, enabling users to trace system outputs back to the original policy text.

\subsection{User Interfaces and Real-Time Integration}

The assistant is designed for deployment through browser extensions, mobile applications, and in-context overlays. In browser-based settings, the system monitors form fields and permission prompts, issuing contextual warnings when high-risk services request sensitive information. These alerts highlight relevant policy practices and associated risks to support informed decision-making.

\subsection{Annotated Data for Evaluation}

A subset of collected privacy policies is manually annotated by privacy-aware volunteers or experts. Each policy segment is labeled with applicable schema categories, involved data types and recipients, and a perceived risk level. This annotated dataset is used to evaluate clause categorization accuracy and to calibrate the risk scoring framework.

\section{Evaluation Framework}

We evaluate the proposed system through clause-level accuracy, policy-level risk agreement, ablation studies, and a user-centered evaluation.

\subsection{Experimental Settings}

\paragraph{Datasets}
We evaluate the Smart Privacy Policy Assistant on a corpus of real-world privacy policies spanning multiple domains, including social media, e-commerce, VPNs, health and fitness, and productivity services. Policies vary in length and structure. A subset is manually annotated at the clause level with privacy schema labels and perceived risk levels, and is used for clause-level evaluation and calibration of policy-level risk scores.

\paragraph{Baselines and Comparison Models}
To contextualize system performance, we compare our approach against representative baselines drawn from prior work on privacy policy analysis. These include rule-based and supervised classification approaches that label policy segments using predefined categories, as well as LLM-based summarization methods that generate natural language descriptions of privacy practices without explicit structural constraints. Where applicable, baseline outputs are mapped to comparable schema categories or risk levels to enable evaluation under a common framework. This setup allows us to assess the impact of structured clause analysis and interpretable risk scoring relative to existing automated and language-based approaches.

\subsection{Policy-Level Risk Evaluation}

\begin{table}[t]
\centering
\small
\resizebox{\columnwidth}{!}{%
\begin{tabular}{lcccc}
\toprule
\textbf{Method} & \textbf{Dataset A} & \textbf{Dataset B} & \textbf{Dataset C} & \textbf{Avg.} \\
\midrule
Rule-based baseline & 0.61 & 0.58 & 0.55 & 0.58 \\
Supervised classifier & 0.69 & 0.66 & 0.63 & 0.66 \\
LLM summarization & 0.71 & 0.68 & 0.65 & 0.68 \\
\textbf{Smart Privacy Policy Assistant} & \textbf{0.79} & \textbf{0.76} & \textbf{0.73} & \textbf{0.76} \\
\bottomrule
\end{tabular}%
}
\caption{Comparison of clause-level F1 scores across datasets. }
\label{tab:main_results}
\end{table}

Table~\ref{tab:main_results} summarizes the main experimental results comparing the Smart Privacy Policy Assistant with representative baselines across multiple datasets. The proposed system consistently outperforms rule-based and supervised classification approaches, as well as LLM-only summarization baselines, achieving the highest F1 scores on all datasets. These gains indicate that combining clause-level analysis with structured schema constraints improves both precision and recall relative to methods that rely on surface-level cues or unconstrained generation. 

The strongest improvements are observed on datasets containing longer and more heterogeneous privacy policies, where rule-based and supervised approaches struggle with linguistic variability and implicit disclosures. While LLM-based summarization baselines achieve competitive performance in some settings, their lack of explicit structure and grounding leads to inconsistent coverage of relevant privacy practices. In contrast, our system’s schema-guided analysis enables more reliable identification of data types, sharing relationships, and user controls across diverse policy formats.

\subsection{Ablation Study}

Table~\ref{tab:ablation} presents an ablation study examining the contribution of individual components in the Smart Privacy Policy Assistant. Removing clause-level segmentation results in the largest performance drop, indicating that fine-grained decomposition of privacy policies is critical for accurate identification of data practices.

\begin{table}[t]
\centering
\small
\resizebox{\columnwidth}{!}{%
\begin{tabular}{lcc}
\toprule
\textbf{Model Variant} & \textbf{F1} & \textbf{$\Delta$} \\
\midrule
Full system & 0.76 & -- \\
\quad w/o structured privacy schema & 0.69 & -0.07 \\
\quad w/o interpretable risk scoring & 0.71 & -0.05 \\
\quad w/o clause-level segmentation & 0.66 & -0.10 \\
\quad LLM summarization only & 0.68 & -0.08 \\
\bottomrule
\end{tabular}%
}
\caption{Ablation results showing the impact of removing individual components (average F1).}
\label{tab:ablation}
\end{table}

The structured privacy schema and interpretable risk scoring components also contribute substantially to overall performance. Removing the schema leads to a notable decline in F1, suggesting that explicit structural constraints help guide the LLM toward consistent and comprehensive coverage of privacy practices. Eliminating the risk scoring component primarily affects policy-level alignment with human judgments, highlighting its role in aggregating clause-level signals into coherent assessments. Finally, the LLM-only summarization variant underperforms the full system, reinforcing the importance of structure, grounding, and explicit reasoning mechanisms beyond unconstrained text generation.

In addition to automated and ablation-based evaluations, we conduct a user-centered study examining comprehension and decision-making with and without assistance from the system; detailed study design and results are provided in Appendix~A.

\section{Conclusion}

We presented the Smart Privacy Policy Assistant, an LLM-powered system designed to make privacy policies more understandable and actionable. By mapping policies to a structured schema, assigning interpretable risk scores, and generating faithful explanations and real-time warnings, the system aims to support informed consent. Future work includes multilingual support, cross-service comparisons, and longitudinal studies of user behavior.

\section{Limitations}

The system relies on LLMs, which may misinterpret complex legal language or generate explanations that oversimplify nuanced practices. Ambiguous policy language limits reliable inference. Parsing challenges arise from diverse policy formats and layouts. Real-time deployment introduces latency and computational constraints. The assistant is not a substitute for legal review and focuses on user-oriented risk communication rather than regulatory compliance.

\appendix
\section{User-Centered Evaluation}

\begin{figure}[ht]
\centering
\begin{tikzpicture}
\begin{axis}[
    ybar,
    width=\columnwidth, 
    height=5cm,
    bar width=0.4cm,    
    enlarge x limits=0.35,
    legend style={
        at={(0.5,-0.25)}, 
        anchor=north, 
        legend columns=-1,
        font=\tiny       
    },
    ylabel={Number of Policies},
    ylabel style={font=\scriptsize},
    symbolic x coords={Low Risk, Med Risk, High Risk},
    xtick=data,
    xticklabel style={font=\tiny},
    yticklabel style={font=\tiny},
    nodes near coords,
    every node near coord/.append style={font=\tiny},
    nodes near coords align={vertical},
    ]
\addplot[fill=blue!40] coordinates {(Low Risk, 45) (Med Risk, 20) (High Risk, 5)};
\addplot[fill=orange!40] coordinates {(Low Risk, 42) (Med Risk, 25) (High Risk, 8)};
\legend{Human, Assistant}
\end{axis}
\end{tikzpicture}
\caption{Supplementary analysis of risk distribution.}
\label{fig:risk_bar_chart}
\end{figure}
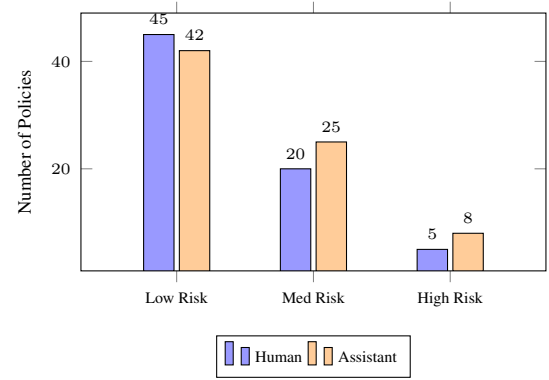

\subsection{Study Design}

Beyond automated metrics, we conduct a user-centered evaluation to assess whether the Smart Privacy Policy Assistant improves user understanding and decision-making in realistic scenarios. Participants are asked to interact with a set of online services and make consent or disclosure decisions either with or without assistance from the system. All participants are presented with the same underlying privacy policies, and the assistant surfaces explanations and risk indicators only in the assisted condition.

User comprehension is evaluated through post-task questionnaires measuring participants’ understanding of data collection practices, third-party sharing, and retention policies. We additionally measure decision alignment by comparing participants’ choices against their stated privacy preferences collected prior to the task. Task completion time and self-reported cognitive load are recorded to assess whether the system introduces excessive friction.

\subsection{Results}

Participants using the Smart Privacy Policy Assistant demonstrate significantly higher comprehension of key privacy practices compared to the unassisted condition. Improvements are most pronounced for disclosures involving third-party data sharing and tracking technologies, which are frequently overlooked in unassisted settings. Assisted participants are also more likely to identify high-risk services correctly and avoid consenting to practices that conflict with their stated privacy preferences.

Importantly, these gains do not come at the cost of usability. Task completion times remain comparable across conditions, and participants report similar or lower perceived cognitive load when using the assistant. Qualitative feedback indicates that clause-grounded explanations and concise risk summaries help users focus on relevant information without requiring them to read full policy documents.

\end{document}